\documentstyle[aps,epsfig,prl,floats]{revtex}
\newcommand{\be}{\begin{equation}}
\newcommand{\ee}{\end{equation}}
\newcommand{\bea}{\begin{eqnarray}}
\newcommand{\eea}{\end{eqnarray}}

\begin{document}
%\draft
\title{Series study of the One-dimensional S-T Spin-Orbital Model}
\author{
        Weihong Zheng\cite{zwh} and J. Oitmaa\cite{otja}
}
\address{School of Physics, University of New South Wales, Sydney NSW
2052, Australia
}
%@twocolumnfalse\endcsname
\date{\today}
\maketitle
\widetext

\begin{abstract}
We use perturbative series expansions about a staggered dimerized ground 
state to compute the ground state energy, triplet excitation spectra and 
spectral weight for a one-dimensional model 
in which each site has an
$S=\case 1/2$ spin ${\bf S}_i$ and a pseudospin ${\bf T}_i$, representing
a doubly degenerate orbital. An explicit dimerization is introduced to 
allow study of the confinement of spinon excitations. The elementary
triplet represents a bound state of two spinons, and is stable over 
much of the Brillouine zone. A special line is found in the gapped spin-liquid
phase, on which the triplet excitation is dispersionless. The formation of
triplet bound states is also investigated.

\end{abstract}


%]
%\narrowtext

\section{Introduction}
There has been considerable interest recently in magnetic systems with 
orbital degeneracy\cite{kug82,fei97,ish97,pat98,miz00,kha00}.
Interesting physics in such systems includes the possibility of
orbital ordering and the influence of this on magnetic
order, the existence and properties of coupled excitations involving
both spin and orbital degrees of freedom, and the relation to metal-insulator
transitions. Materials in which such effects are believed to manifest
themselves include the perovskite KCuF$_3$\cite{fei97},
pure and doped manganites La$_{1-x}$Sr$_x$MnO$_3$\cite{ish97,miz00},
Na$_2$Ti$_2$Sb$_2$O and NaV$_2$O$_5$\cite{pat98}, LaTiO$_3$\cite{kha00},
and others.

Realistic theoretical models of such
materials must start from a rather complex Hubbard type
model (see e.g. \cite{miz00}) with many parameters.
To avoid these complications many authors consider an
appropriate strong coupling limit and electron density so that
the charge degrees of freedom are effectively frozen out.
The orbital state, for a system with two degenerate orbitals,
can then be represented by a pseudospin ${\bf T}_i$.
Although, in general, the pseudospin interactions will not be
rotationally invariant, this is often ignored.
Thus much recent work has focussed on the ``$S$-$T$ model"
\begin{equation}
H = J_1 \sum_{\langle ij\rangle } {\bf S}_i \cdot {\bf S}_j
+ J_2 \sum_{\langle ij\rangle } {\bf T}_i \cdot {\bf T}_j
+ K \sum_{\langle ij\rangle } ({\bf S}_i \cdot {\bf S}_j) 
	( {\bf T}_i \cdot {\bf T}_j) \label{Horg}
\end{equation}
both in one-dimension\cite{pat98,aza99,yam00,ito,yam,li98,gri99,li99}
and in higher dimensions\cite{jos99,san99,van51}.

In this paper we consider the one-dimensional case which can be written
% , up to a constant, 
in the form
\begin{equation}
H = K \sum_i [ ( {\bf S}_i \cdot {\bf S}_{i+1} + j_2 ) ( {\bf T}_i \cdot {\bf T}_{i+1} + j_1 )
 - j_1 j_2 ]
\end{equation}
where $j_{1,2}\equiv J_{1,2}/K$. The constant $K$ sets the
energy scale and can be set equal to 1. Thus we are left with a 
2-parameter Hamiltonian.

It is  known that, at the point $j_1=j_2=1/4$, the Hamiltonian has SU(4)
symmetry, and is exactly solvable, by the Bethe ansatz method\cite{sut75,li98,li99}.
At this point the model is gapless with central charge $c=3$
and 3 bosonic modes. At the point $j_1=j_2=3/4$ the model has an exact ground
state\cite{kol98} in which spin and orbital operators form dimerized singlets
in a staggered pattern. This observation is based on an equivalent realization
of the model as a 2-leg ladder coupled by a 4-spin interaction\cite{ner97}.
This ``matrix product state" is doubly degenerate and gapped. The phase diagram
of the model thus contains both gapped and gapless regions. The various
$T=0$ phases and their boundaries have been mapped out by
Itoi {\it et al.}\cite{ito}, using the DMRG technique.

A number of interesting questions can also be posed regarding the low energy
excitations, both in gapless and gapped regions. It is by now well
known that the elementary excitations in a single spin-half
antiferromagnetic chain are $S=\case 1/2$ spinons\cite{Faddev:81}. In a system with periodic
boundary conditions spinons are formed in pairs, and excitations
form a two-spinon continuum. However addition of dimerization to a single chain, 
or coupling of two chains to form a ladder leads to confinement
of the spinons into bound $S=1$ (triplet or magnon) excitations. 
Spontaneous dimerization, as in the Majumdar-Ghosh model\cite{MG},
 leads to two degenerate ground states.
% A dimerized spin chain, either explicit in the Hamiltonian or formed
% spontaneously, as in the Majumdar-Ghosh model\cite{MG}, has two degenerate
% ground states. 
In such a case the elementary excitation is a domain wall, 
separating the two ground states. This excitation also has $S=\case 1/2$ and
is also refered to as a spinon. There has been much recent study of these issues,
including the dynamical transition from spinons to 
triplets/magnons\cite{sor98,uhr99,oleg99,zwhj12d}.

In this paper we consider a generalized $S$-$T$ model in which staggered dimerization is 
introduced explicitly into the $S$ and $T$ chains, as shown in Figure 1(a).
There are two reasons for doing this. The first is to provide a
natural separation of the Hamiltonian into an unperturbed
part and a perturbation, allowing the development of ``dimer" series
expansions\cite{gel00}. Secondly, by varying the dimerization
parameter we are able to move from regions of integer spin excitations to
$S=\case 1/2$ excitations, and study the dynamical transition from triplet to spinon
excitations in this model.

We study the ground state energy, energy gap, and excitation spectrum
for general values of $(j_1,j_2)$ within
the gapped phase. 
At the point $j_1=j_2=\case 3/4$ the system has an exact dimer ground state, as
shown by Kolezhuk and  Mikeska\cite{kol98}. We refer to this as the K-M point. The elementary
excitations will then be spinons. We would expect spontaneous dimerization to exist
away from this point, perhaps throughout the gapped phase. However when $x<1$
one dimerized state becomes the unique ground state and the
spinons will become confined into $S=1$ triplet excitations. At the
K-M point the dimerized state is, in fact, an exact ground state for any $x$.
At this point we are able to compute substantially longer series for the triplet
dispersion and spectral weight. This allows us to study the transition from triplet to
deconfined spinons as dimerization $\delta = 1-x$ vanishes.

We compute dispersion curves for triplet excitations away from the
K-M points, specifically along the lines $j_1=j_2$, and for $j_2=1$. In each
case there is a parameter value where the dispersion curve is flat, and
we identify a special line in the $(j_1,j_2)$ plane having dispersionless
excitations. This must mean that along this line excitations are
localized objects which do not propagate.

\section{Dimer Series Expansions}

\begin{figure}
\begin{center} 
\vskip 13mm
\epsfig{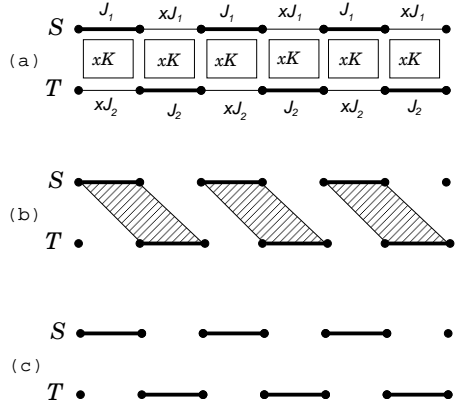}
\caption[]
          {(a) The  dimerized $S$-$T$ chain, showing the various interaction terms;
          (b) decoupled plaquettes forming the unperturbed system;
          (c) one of two exact ground states at the Kolezhuk-Mikeska point. The
          solid lines represent $S=0$ singlet pairs.}
  \label{fig_1} 
  \end{center}   
 \end{figure}

In this paper, we study the generalized version of the $S$-$T$  model, namely the staggered dimerized
spin-orbit model,  using the dimer series expansion approach\cite{gel00}.
% Our calculations are based on dimer series expansions\cite{zwh,gel00} about the
% exact KM gound state. 
It is convenient, for this purpose, to view the
system as a 2-leg ladder, as shown in Figure 1(a).
The Hamiltonian for it can be split as
\begin{equation}
H=H_0 + x V
\end{equation}
where 
\begin{equation}
H_0 = J_1 \sum_{i} {\bf S}_{2i-1} \cdot {\bf S}_{2i} + 
J_2 \sum_{i} {\bf T}_{2i} \cdot {\bf T}_{2i+1}
\end{equation}
and
\begin{equation}
V = J_1 \sum_{i} {\bf S}_{2i} \cdot {\bf S}_{2i+1} + 
J_2 \sum_{i} {\bf T}_{2i-1} \cdot {\bf T}_{2i}
+ K \sum_{i} ( {\bf S}_{i} \cdot {\bf S}_{i+1}) ({\bf T}_{i} \cdot {\bf T}_{i+1}) \label{eqV}
\end{equation}
Here $\delta=1-x$ is the dimerization parameter, and when $x\to 1$, 
the dimerization become zero, and
one recovers the original Hamiltonian in Eq. \ref{Horg}. 
Our strategy is then to expand the quantities of interest in powers 
of $x$ and to use standard analysis methods\cite{gut} to evaluate these series 
at $x=1$.
The unperturbed Hamiltonian $H_0$ consists of decoupled plaquettes, as shown in Figure 1(b).
The ground state of each plaquette is a product of $S$ and $T$ singlets on 
the respective bonds and the ground state of $H_0$ is then a direct product over plaquettes.
This state is shown schematically in Figure 1(c), and has
energy $E_0 = -3 (J_1+J_2) N/16$, where $N$ is nnumber of spins (including pseudospin). 
It remains an exact ground state of
the full Hamiltonian at the Kolezhuk-Mikeska (K-M) point $j_1=j_2=3/4$.
The perturbation $V$ (Eq. \ref{eqV}) includes all interactions between plaquettes.

To see that the staggered dimer state is the ground state of $H$ at $j_1=j_2=3/4$
(for any $x$) we write $V$ as
\begin{equation}
V =  \sum_{i} {\bf S}_{2i} \cdot {\bf S}_{2i+1} (J_1 + K {\bf T}_{2i} \cdot {\bf T}_{2i+1} )
+ \sum_{i} {\bf T}_{2i-1} \cdot {\bf T}_{2i} ( J_2 + K {\bf S}_{2i-1} \cdot {\bf S}_{2i} )
\end{equation}
Each of the operators $(3/4 + {\bf T}_{2i} \cdot {\bf T}_{2i+1})$,
$(3/4 + {\bf S}_{2i-1} \cdot {\bf S}_{2i})$ gives zero when acting
on this state and hence it is an eigenstate of $V$ with eigenvalue zero,
and thus an eigenstate of the full Hamiltonian.

\section{Results}

\subsection{General $(j_1, j_2)$}

We have obtained series to order $x^{11}$ for the
ground state energy and to order $x^9$ for the elementary triplet excitation energy.
The latter corresponds to excitation of either an $S$ or $T$ singlet dimer to
an $S$ or $T$ triplet, followed by propagation of this triplet along the ladder. We  do not
include the series coefficients here but can provide them on request.

\begin{figure}
\begin{center} 
\vskip -13mm
\epsfig{file=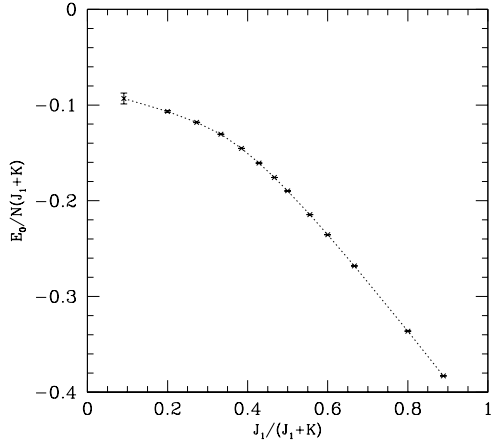, width=8cm}
\vskip -10mm
\caption[]
          {The rescaled ground state per site  with $x=1$, $j_1=j_2$.}
  \label{fig_e0} 
  \end{center}   
 \end{figure}

The series are then evaluated at the point $x=1$ using  integrated
differential approximants\cite{gut}. Figure \ref{fig_e0} shows the ground state energy
along the line $j_1=j_2$, and includes the exact SU(4) and K-M
values. The series converges well, even into the gapless region
$j_1=j_2<1/4$, and the energy varies smoothly, consistent with the expected
Kosterlitz-Thouless  %second-order 
transition at the SU(4) point.

\begin{figure}
\begin{center} 
\vskip -13mm
\epsfig{file=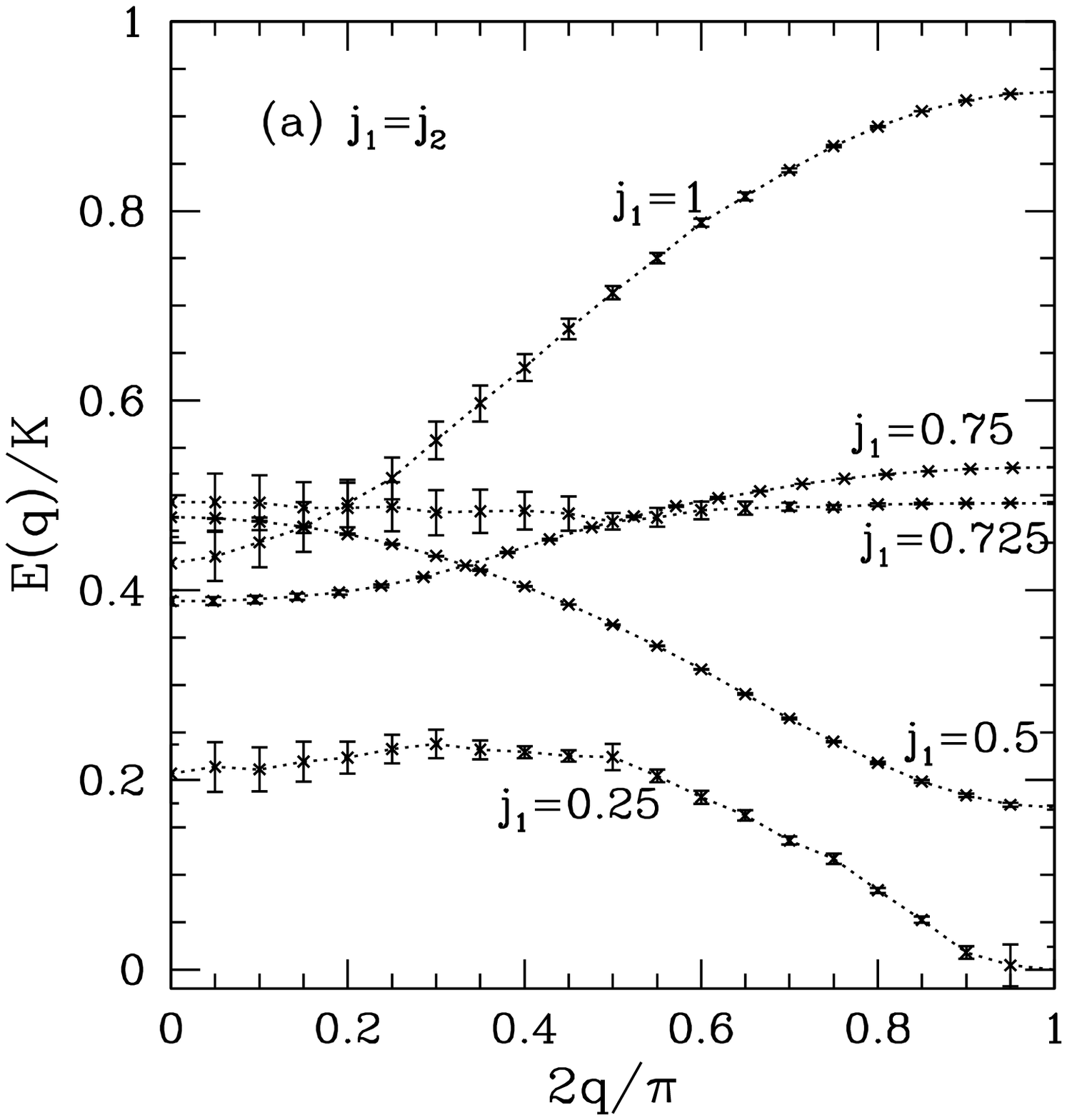, width=8cm}
\epsfig{file=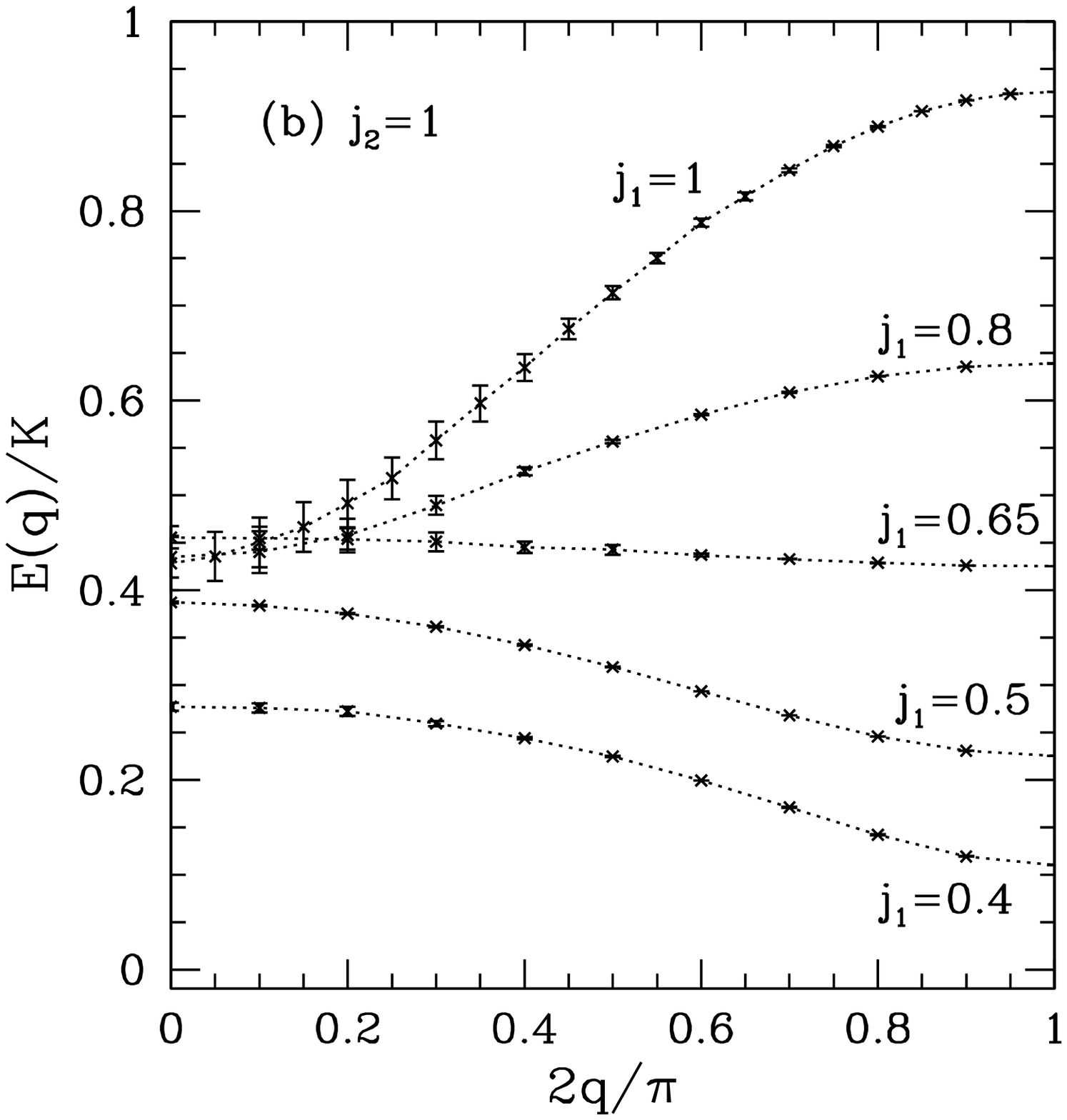, width=8cm}
\vskip -10mm
\caption[]
          {The elementary triplet excitation spectra along the 
          line $j_1=j_2$ (a) and the line $j_2=1$ (b) for the system with $x=1$.}
  \label{fig_mk} 
  \end{center}   
 \end{figure}

Figure \ref{fig_mk}  show triplet excitation energies versus 
wavevector $q$ for
the case of zero dimerization $(x=1)$ firstly along the line $j_1=j_2$,
and secondly along the line $j_2=1$. In both cases there is a qualitative change
in the dispersion curves, with the minimum
energy at $q=\pi/2$ for small $j$ moving to $q=0$ for large $j$. 
In each case we identify a critical
$j_1$ for which the dispersion curve is virtually flat. In figure \ref{fig_mk}(a)  
this occurs at approximately $(j_1, j_2)=(0.725, 0.725)$,
slightly lower than the K-M point, whereas in figure \ref{fig_mk}(b) the
dispersion curve is flat at $(j_1,j_2)\simeq (0.65, 1)$.
Figure \ref{fig_mk}(a) shows the excitation energy at $q=\pi/2$ continuously
going to zero as $j_1=j_2$ reduces to 0.25, the
SU(4) point.

\begin{figure}
\begin{center} 
\vskip -13mm
\epsfig{file=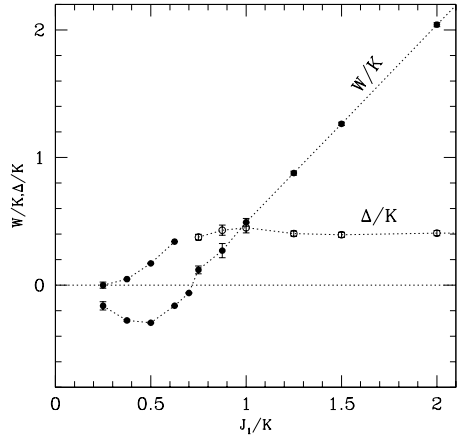, width=8cm}
\vskip -10mm
\caption[]
          {The minimum gap and bandwidth for elementary triplet 
          excitations for  $x=1$, $j_1=j_2$.}
  \label{fig_gap_W} 
  \end{center}   
 \end{figure}

From the dispersion curves we can compute the minimum gap $\Delta$,
and the bandwidth $W=E(\pi/2)-E(0)$. These are shown in Figure \ref{fig_gap_W} for $j_1=j_2$ (and 
$x=1$). We see that the gap opens at $j_1=j_2=0.25$, increases monotonically
to the K-M point, and remains essentially constant (i.e. $\Delta \propto K$)
thereafter. The bandwith $W$ is zero when the dispersion is flat. Negative 
bandwidth by our convention means $E(q=\pi/2) < E(q=0)$, for larger $j_1$,
the bandwidth $W$ is proportional to $J_1$.

\begin{figure}
\begin{center} 
\vskip -13mm
\epsfig{file=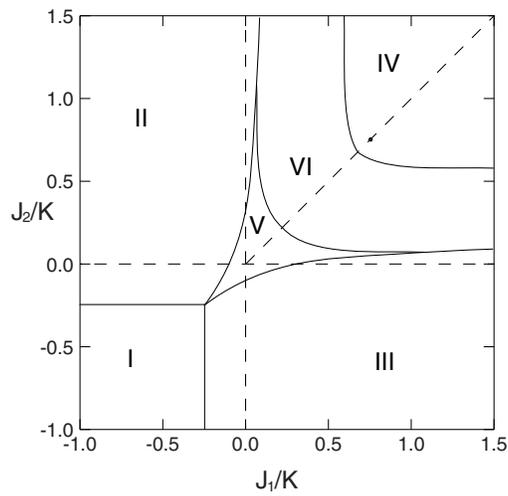, width=8cm}
\vskip -10mm
\caption[]
          {The phase diagram for system with $x=1$ (after ref. \onlinecite{ito}.}
  \label{fig_pd} 
  \end{center}   
 \end{figure}

Itoi {\it et al.}\cite{ito} have displayed a ground state phase diagram for the 
model, and designated various phases by symbols I-IV. We reproduce this 
diagram in Figure \ref{fig_pd}, where we have separated the gapped phase IV into two parts
IV and VI. The line dividing the regions is where the triplet excitation is
dispersionless.

\subsection{$(j_1,j_2)=(3/4,3/4)$}
As mentioned above the staggered dimerized singlet state is the exact ground state 
for any value of the dimerization $\delta=1-x$. This feature makes it possible to 
derive much longer series in $x$. This feature was also observed along the
Shastry-Sutherland line in the $J_1-J_2-\delta$ chain\cite{zwhj12d,j12d2p}; 
we do not have a simple
physical argument to explain it.
We have computed series to order $x^{25}$ for both the triplet excitation spectrum and
the spectral weight.

The primary focus here is to investigate the transition
from triplet to spinon excitations (deconfinement of
spinons) as $x\to 1$. In the disconnected dimer limit $x=0$ the elementary excitations
are localized triplets on one of the spin pairs 
$({\bf S}_{2i-1} \cdot {\bf S}_{2i})$ or $({\bf T}_{2i} \cdot {\bf T}_{2i+1})$.
These triplet excitations develop a dispersion for $x\not= 0$ but remain well
defined for all $x<1$. At $x=1$ the Hamiltonian has full
translation symmetry but, as in the well-known 
Majumdar-Ghosh model\cite{MG}, this symmetry is spontaneously broken leading to 
two degenerate ground states, one of which remains the ground state for $x<1$.
In this case the elementary excitations are domain walls with $S=\case 1/2$
interpolating between the degenerate ground states.

\begin{figure}
\begin{center} 
\vskip -13mm
\epsfig{file=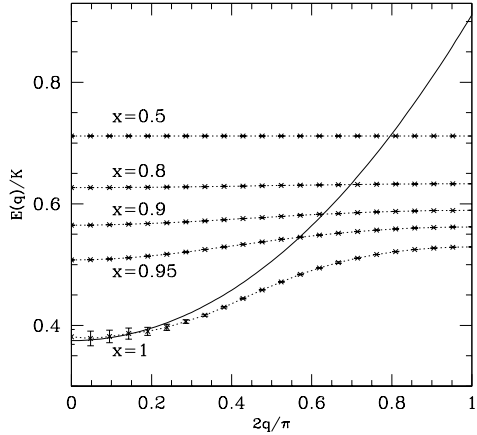, width=8cm}
\caption[]
          {The triplet dispersion relation for system with 
          $j_1=j_2=3/4$ and various $x$. The solid curve
          are the lower edge of the two-spinon continuum for $x=1$,
given by Kolezhuk and  Mikeska\cite{kol98}.}
  \label{fig_mk_yp75} 
  \end{center}   
 \end{figure}

Following Singh and Zheng\cite{zwhj12d} we start from the small $x$ limit,
and look for breakup of the triplet excitations as $x\to 1$. The triplet 
dispersion relation is shown in Figure \ref{fig_mk_yp75} for various values of $x$.
The curves are quite flat for small $x$, and start to develop some dispersion by $x\sim 0.9$.
We expect that the gap at $q=0$ behaves as
\be
\Delta \sim \Delta_0 + c (1-x)^{\nu}
\ee
but the series are not regular enough to confirm this or to estimate the exponent $\nu$.
The lower edge of the two-soliton/spinon continuum for $x=1$,
given by Kolezhuk and  Mikeska\cite{kol98} is shown as a solid
line in Figure \ref{fig_mk_yp75}. In the small $q$ region our triplet dispersion agrees
well with this curve, indicating that there are no soliton bound states at small $q$.
However for most of the range of $q$ our triplet dispersion is substantially below the
two-soliton continuum, indicating the formation of soliton
bound states. 
Kolezhuk and  Mikeska\cite{kol98} give a variational ansatz for a triplet excitation 
representing a spinon bound state. Our triplet excitation has signficantly
lower energy, and is the favoured triplet state.
%Our triplet energy differs significantly from the bound
%state shown in reference \onlinecite{kol98}.

\begin{figure}
\begin{center} 
\vskip -13mm
\epsfig{file=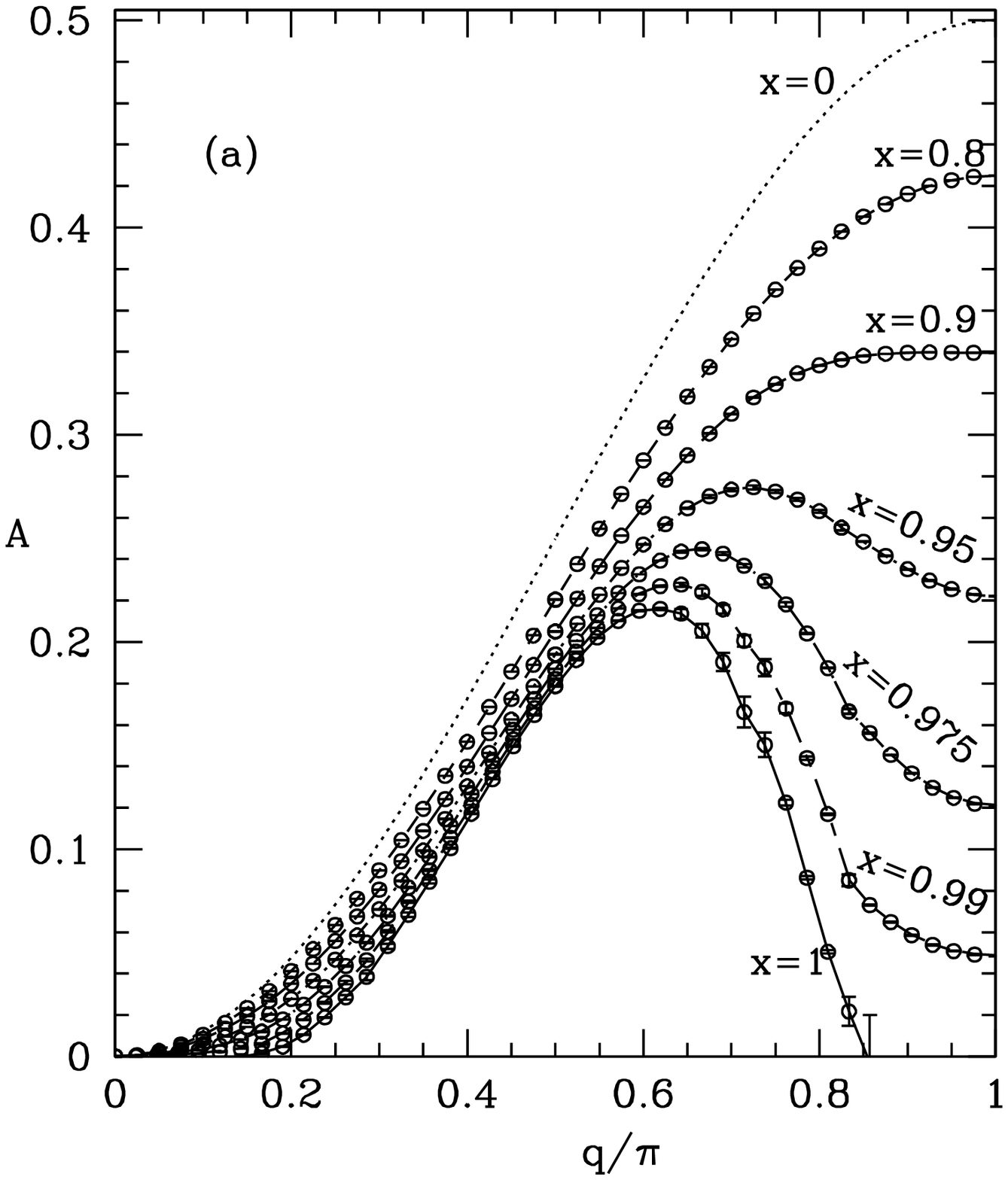, width=8cm}
\epsfig{file=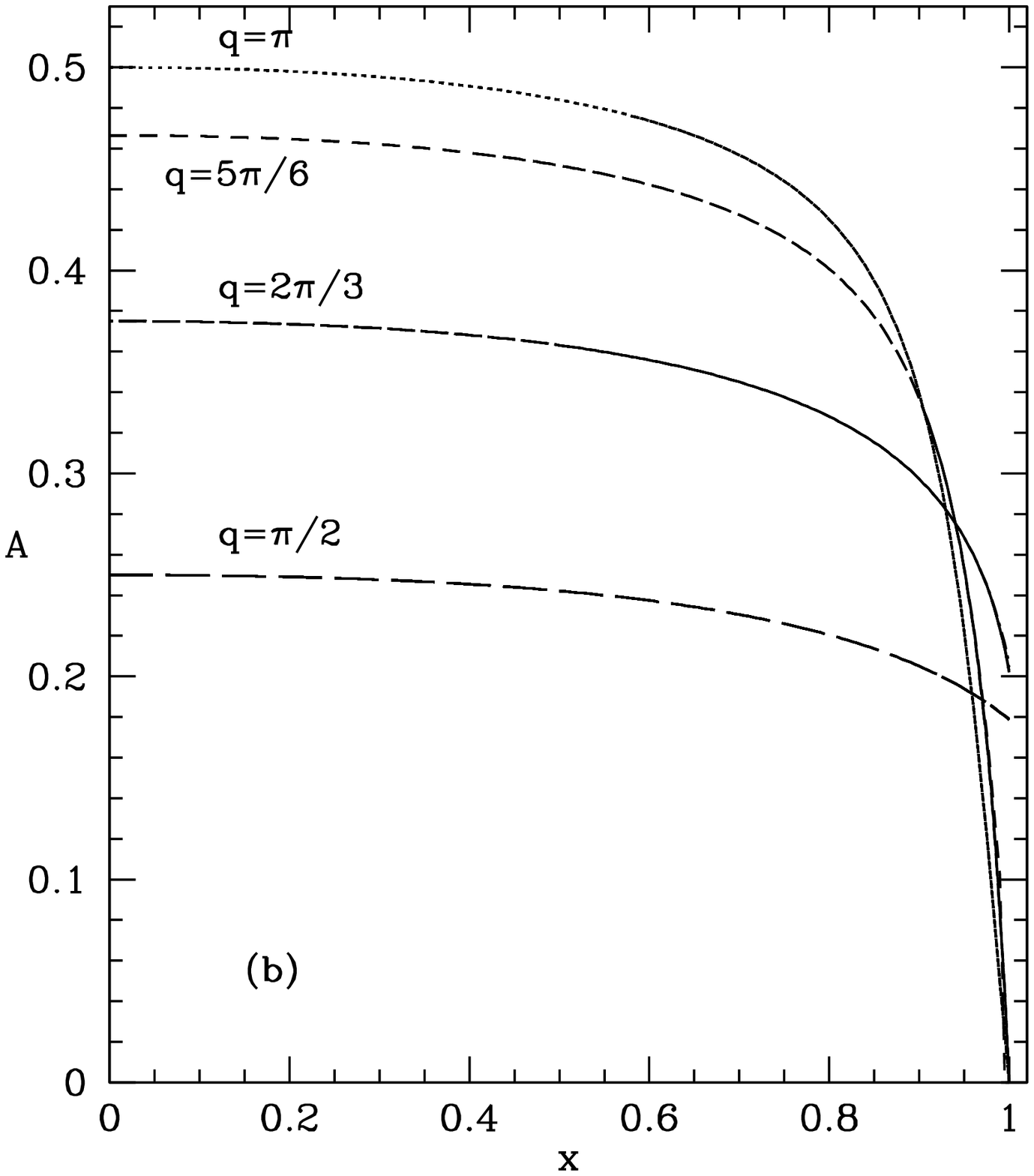, width=8cm}
\vskip -10mm
\caption[]
          {(a). The spectral weight $A$ versus momentum $q$
          for the system along the  line $j_1=j_2=3/4$
for various $x$ (as shown).
(b) The spectral weight $A$ versus  $x$
          for the system along the line $j_1=j_2=3/4$
and various momentum $q$.
}
  \label{fig_Ak} 
  \end{center}   
 \end{figure}

We have also computed series to order $x^{25}$ for the spectral weight $A(q)$, and
we show results for various $x$ values in Figure \ref{fig_Ak}. 
For this model
the spectral weight satisfies the relation
\begin{equation}
A(q)(1-\cos{(\pi-q)})=A(\pi-q)(1-\cos{q}).
\end{equation}
Figure \ref{fig_Ak}(a) shows a dramatic change in the spectral weight as $x\to 1$, approaching the
transition to deconfined spinons. At $x=1$ the spectral weight remain finite 
for $q_c < q < \pi-q_c$, but vanishes for $q<q_c$ or $q>\pi - q_c$. The
critical wavenumber is estimated to be about $q_c \simeq \pi/6$, and
this agrees well the triplet dispersion results shown in Figure \ref{fig_mk_yp75}.
The region over which $A(q)$ remains finite is substantially larger than that for the
$J_1-J_2-\delta$ chain\cite{zwhj12d}, indicating that spinon bound states are
more stable in the present model. A simple Dlog Pad\'e
analysis of the spectral weight $A(q)$ for $q<q_c$ or $q>\pi - q_c$
suggests that $A(q)$ vanishes linearly with $(1-x)$, in agreement with the
analysis in ref. \onlinecite{oleg99}. This is shown in Figure  \ref{fig_Ak}(b)
where we show $A(q)$ as a function of $x$ for various $q$ values.

\subsection{Two-Triplet bound states}

\begin{figure}
\begin{center} 
\vskip -13mm
\epsfig{file=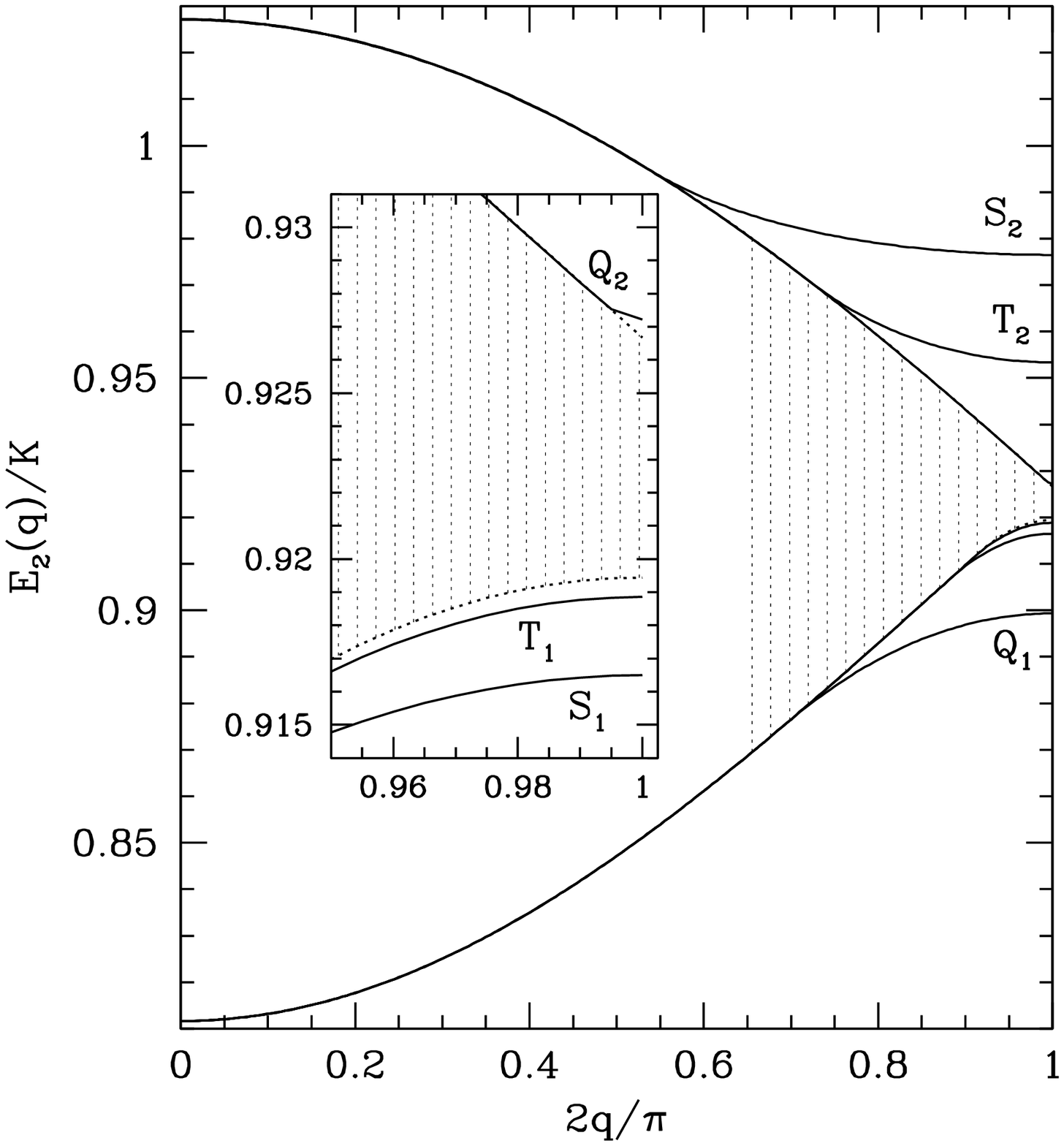, width=8cm}
\vskip -10mm
\caption[]
          {
          The two-particle excitation spectrum of the 
          staggered dimerized spin-orbit chain with
          $j_1=0.5, j_2=0.75$ and $x=0.5$. Beside the two-particle
          continuum (gray shaded), there is
          one singlet bound state ($S_1$),
          one triplet bound state ($T_1$), and one quintet bound state
          ($Q_1$) below the continuum, and one singlet antibound state ($S_2$),
          one triplet antibound state ($T_2$), and one
          quintet antibound state ($Q_2$) above the continuum.
          The inset enlarges the region near 
          $k=\pi/2$ so we can see $S_1$, $T_1$ and $Q_2$ below/above the continuum.
          }
  \label{fig_mk_2p} 
  \end{center}   
 \end{figure}

It is also of interest to investigate the possible existence of bound states of the
elementary triplet excitations. We have done this, using a recently developed series
techniques for multiparticle excitation\cite{zwh2p}. 
The series for two-triplet bound states have been computed up to order $x^6$ for singlet
bound states, and to order $x^9$ for triplet and quintet states.
The reason why the singlet series is computed to only 6th order compared
to 9th order for the triplet and quintet states is that the
singlet has the same quantum numbers as the ground state. Thus
a much more elaborate orthogonalization method is required to implement the
cluster expansion for the singlet. 
Figure \ref{fig_mk_2p} shows
the excitation spectrum of the staggered dimerized $S$-$T$ chain for parameters
$(j_1,j_2)=(0.5,0.75)$ and dimerization parameter $\delta=1-x=0.5$.
For $j_1<j_2$ (as here) the $S$ triplets will have lowest energies, and the
figure refers to these. The figure show a two-triplet continuum, with a number of bound 
states below the continuum, near $q=\pi/2$ as well as ``antibound"
states above the continuum. The lowest bound state is an $S=2$
quintet, with $S=0$ singlet and $S=1$ triplet bound states at higher energy.
The behaviour
is qualitatively similar to the case of $J_1-J_2-\delta$ chain\cite{j12d2p},
but the quantum numbers of the bound states differ.

\begin{figure}
 \begin{center} 
\vskip -0.5cm
\epsfig{file=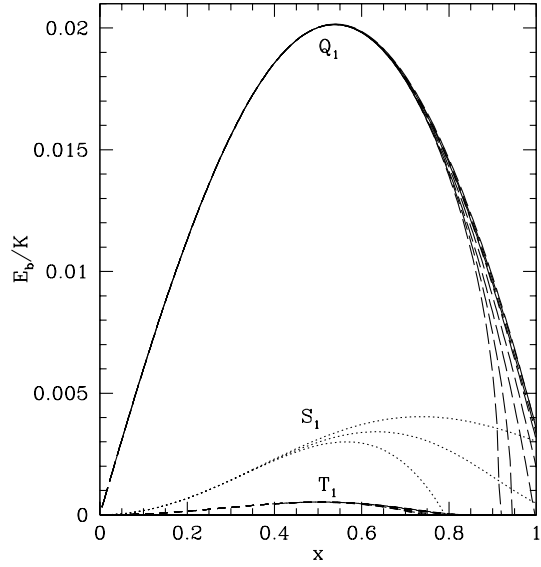, height=10cm}
\vskip -1.2cm
 \caption[]
          {The binding  energy $E_b/K$ at $q=\pi/2$ versus dimerization 
          $x$ for  singlet, triplet, and quintet  bound states  ($S_1$, $T_1$ and $Q_1$)
          of the staggered dimerized spin-orbit chain with $j_1=0.5$ and $j_2=0.75$.  
         Several different integrated differential approximants to
   the series are shown.}
 \label{fig_Eb_STQ} 
 \end{center}
\end{figure}

To investigate the binding energy as a function of
the dimerization parameter we have repeated this calculation for different values 
of $x$, and show the results in Figure \ref{fig_Eb_STQ},
as a binding energy measured relative to the lower edge of the 
continuum. We show results for $q=\pi/2$ only, as at this point the
binding energy is largest. The binding energy vanishes at $x=0$, as
expected, since the triplet excitations there do not
interact. 
The $Q_1$ bound state energy vanishes linearly at small $x$, whereas
the $S_1$ and $T_1$ states vanish quadratically. The binding energies
are greatest for $x$ around 0.5, and vanish as $x\to 1$.

\section{Discussion and Conclusions}
We have used series expansion methods to investigate properties of a spin model,
the dimerized $S$-$T$ chain, which has received much attention recently in
connection with orbital ordering in magnetic  systems. By including a
staggered dimerization in the $S$ and $T$ exchange interactions we are able to derive
perturbation expansions about a state which is the exact dimerized ground
state at the Kolezhuk-Mikeska point $(j_1,j_2)=(3/4,3/4)$, and
these expansions are regular and analysable throughout the gapless region of
the model's phase diagram.

As well as the ground state energy, we have obtained dispersion curves for elementary
triplet excitations. The latter show an interesting feature, where
the minimum excitation energy changes from $q=\pi/2$ to
$q=0$ as $j_1,j_2$ increase, and the crossover is marked by dispersionless excitations
along a particular line in the phase diagram, which we locate.
The K-M point lies close to this line, but not on it. At the K-M point,
using longer series, we find a definite though small dispersion.

In the case of zero dimerization, in the gapped region of phase diagram, 
it might be 
supposed that elementary
excitations consist of deconfined spinons (domain walls). However we
show that triplet excitations have lower energy than that of two spinons, and
hence that spinons will bind, except for low momentum. 

We have also used a recently developed series technique to study
2-triplet excitations, and find a variety of bound states below
and above the continuum. 

After completing this work we learnt of the
paper by Martins and Nienhuis\cite{mar00}. They treat exactly
the same model as discussed here, but for the case $K<0$. They find two points
at which the model is exactly solvable: one at $j_1=j_2=-1/4$, $x=1$, 
the other at $j_1=j_2=-1/2$, $x=1/2$. These lie within phase I (Fig. \ref{fig_pd})
where the ground state is fully ferromagnetic.

\acknowledgments
This work forms part of a research project supported by the
Australian Research Council. We thank Prof. Oleg Sushkov for useful discussions.
The computation has been performed on
Silicon Graphics Power Challenge.


\begin{references}
\bibitem[*]{zwh} Email address: w.zheng@unsw.edu.au
\bibitem[\dag]{otja} Email address: j.oitmaa@unsw.edu.au

%1:
\bibitem{kug82}K.I. Kugel and D.I. Khomskii, Sov. Phys. Usp. {\bf 25},
231(1982).
%2:
\bibitem{fei97}L.F. Feiner, A.M. Ol\'es \& J. Zaanen, Phys. Rev. Lett.
{\bf 78}, 2799(1997).
%3:
\bibitem{ish97}S. Ishihara, J. Inoue and S. Maekawa, Phys. Rev. B {\bf 55}, 8280(1997).
%4
\bibitem{pat98}S. Pati, R.R.P. Singh and D.I. Khomskii, Phys. Rev. Lett.
{\bf 81}, 5406(1998).
%5:
\bibitem{miz00}T. Mizokawa, D.I. Khomskii and G.A. Sawatzky, 
Phys. Rev. B{\bf 61}, R3776(2000).
%6:
\bibitem{kha00}G. Khaliullin and S. Maeckawa, Phys. Rev. Lett. {\bf 85}, 3950(2000).
%7:
\bibitem{aza99}P. Azaria, A.O. Gogolin, P. Lecheminant and A.A. Nersesyan,
Phys. Rev. Lett. {\bf 83}, 624(1999).
%8
\bibitem{yam00}Y. Yamashita, N. Shibata and K, Ueda, J. Phys. Soc. Japan {\bf 69}, 242(2000).
%9:
\bibitem{ito}C. Itoi, S. Qin and I. Affleck, Phys. Rev. B{\bf 61}, 6747(2000).
%10:
\bibitem{yam}Y. Yamashita, N. Shibata and K. Ueda, Phys. Rev. B{\bf 58}, 9114(1998).
%11:
\bibitem{li98}Y.Q. Li, M. Ma, D.N. Shi and F.C. Zhang, Phys. Rev. Lett. {\bf 81}, 3527(1998).
%12:
\bibitem{gri99}B. Frischmuth, F. Mila and M. Troyer, Phys. Rev. Lett. {\bf 82}, 835(1999).
%13:
\bibitem{li99}Y.Q. Li, M. Ma, D.N. Shi and F.C. Zhang, Phys. Rev. B{\bf 60}, 12781(1999).
%14:
\bibitem{jos99}A. Joshi, M. Ma, F. Mila, D.N. Shi and F.C. Zhang, Phys. Rev. B{\bf 60}, 6584(1999).
%15:
\bibitem{san99}G. Santoro, S. Sorella, L. Guidoni, A. Parola and E. Tosatti,
Phys. Rev. Lett. {\bf 83}, 3065(1999).
%16:
\bibitem{van51}M. van den Bossche, F.C. Zhang and F. Mila, Eur. Phys. J. B{\bf 17}, 367(2000).
%17: 
\bibitem{sut75}B. Sutherland, Phys. Rev. B{\bf 12}, 3795(1975).
%18:
\bibitem{kol98}A.K. Kolezhuk and H.J. Mikeska, Phys. Rev. Lett. {\bf 80}, 2709(1998).
%19:
\bibitem{ner97}A.A. Nersesyan and A.M. Tsvelik, Phys. Rev. Lett. {\bf 78}, 3939(1997).
%21 spinons
\bibitem{Faddev:81}
L.~D.~Faddev and L.~A.~Takhtajan, Phys.~Lett.~A, 375 (1981).

% spinons
\bibitem{MG}
C.K. Majumdar and D.K. Ghosh, J. Math. Phys. {\bf 10}, 1399(1969);
C.K. Majumdar, J. Phys. C{\bf 3}, 911(1970); P.M. van den Brook, Phys.
Lett. {\bf 77A},  261(1980). 

%20:
\bibitem{sor98}E.S. S\o rensen, I. Affleck, D. Augier and D. Poilblanc,
 Phys. Rev. B {\bf 58}, R14701(1998).


\bibitem{uhr99}G.S. Uhrig, F. Sch\"onfeld, M. Laukamp and E. Dagotto, Eur. Phys.
J. B{\bf 7}, 67(1999).

\bibitem{oleg99}
T.~M.~Byrnes, M.~T.~Murphy, and O.~P.~Sushkov,
Phys. Rev. B{\bf 60}, 4057(1999).

\bibitem{zwhj12d}
R.~R.~P.~Singh and W.~H.~Zheng, Phys.~Rev.~B {\bf 59}, 9911 (1999). 

\bibitem{gel00} 
M.~P.~Gelfand and R.~R.~P.~Singh, 
Advances in Phys. {\bf 49}, 93 (2000).

% differential approximants
\bibitem{gut}
A.~J.~Guttmann, in {\it Phase Transitions and Critical Phenomena},
edited by C.~Domb and M.~S.~Green (Academic, New York, 1989), Vol.~13.



%\bibitem{zwh98}W.H. Zheng, V. Kotov and J. Oitmaa, Phys. Rev. B{\bf 57}, 11439(1998).

% single particle excitations in cluster expansion method
%\bibitem{gelfand2}
%M.~P.~Gelfand, Solid State Comm. {\bf 98}, 11 (1996).

% Shastry Sutherland model
% \bibitem{Shastry:81}
% B.~S.~Shastry and B.~Sutherland, Phys.~Rev.~Lett. {\bf 47}, 964 (1981).

% cluster expansion method
% \bibitem{he} 
% H-X.~He, C.~J.~Hamer and J.~Oitmaa, J.~Phys.~A {\bf 23}, 1775 (1990).
% \bibitem{gelfand1} 
% M.~P.~Gelfand, R.~R.~P.~Singh, and D.~A.~Huse, 
% J.~Stat.~Phys. {\bf 59}, 1093 (1990).
% 
\bibitem{j12d2p}
W.~H.~Zheng  C.~J.~Hamer,  R.~R.~P.~Singh, S.~Trebst, and H.~Monien,
cond-mat/0010243, to be published.


\bibitem{zwh2p}
W.~H.~Zheng, C.~J.~Hamer, R.~R.~P.~Singh, S.~Trebst, and H.~Monien,
cond-mat/0010354, to be published.
S.~Trebst, H.~Monien, C.~J.~Hamer, W.~H.~Zheng and  R.~R.~P.~Singh,
Phys. Rev. Lett. {\bf 85}, 4373(2000).


\bibitem{mar00}M.J. Martins and B. Nienhuis. Phys. Rev. Lett. {\bf 85}, 4956(2000).

\end{references}
\end{document}